\documentclass[submission,copyright,creativecommons]{eptcs}
\providecommand{\event}{MARS 2017} 
\usepackage{breakurl}             
\usepackage{underscore}           
\usepackage{listings}
\usepackage{graphicx}
\usepackage{fancyvrb}
\usepackage{latexsym}
\usepackage{amssymb}
\usepackage{float}
\usepackage{multicol}
\newfloat{program}{thp}{lop}
\floatname{program}{Table}

\usepackage{latexsym,multicol,color,pstricks}
\usepackage{listings} 
\usepackage{multirow}
\usepackage{hhline}

\lstdefinelanguage{Uppaal}{ 
basicstyle=\ttfamily, 
keywords={after update,assign,before update,break,case,const,continue,
default,else,enum,for,guard,if,meta,process,progress,return,select,
state,sync,switch,trans,system,while},
keywords={[2]broadcast,bool,clock,chan,commit,init,int,scalar,struct,
typedef,urgent,void}, keywordstyle={[2]\bfseries},
keywords={[3]false,true}, otherkeywords={[3]−>},
morekeywords={[3]−>}, keywordstyle={[3]\bfseries},
comment=[l]{//}, morecomment=[s]{/∗}{∗/}, 
commentstyle=\itshape, 
tabsize=4, 
captionpos=b, 
escapechar=@ 
}
\lstdefinelanguage[GUI]{Uppaal}[]{Uppaal}{ 
keywordstyle={[2]\color{black!50!green}}, 
otherkeywords={−>}, keywordstyle={[3]\color{magenta}},
commentstyle={\color{black!50!red}\itshape}, 
literate={{−−>}{$−−>$}3} 
}
\lstdefinelanguage[LIT]{Uppaal}[GUI]{Uppaal}{ 
literate={{−>}{{$\leadsto$} }2 {−−>}{{$\longrightarrow$} }2
{=}{{$\gets$ }}2 {==}{{$==$}}2 {:=}{{$\gets$ }}2 {<=}{{$\leq$ }}2
{>=}{{$\geq$ }}2 {&&}{{$\land$}}2 {||}{{$\lor$}}2 {<>}{{$\Diamond$}}1
{[]}{{$\Box$}}1 {forall}{{$\forall$}}1 {exists}{{$\exists$}}1}
}

\graphicspath{{images/}}

\title{Evaluating The Stream Control Transmission Protocol Using Uppaal}
\author{Shruti Saini
\institute{School of Computing, Information and Mathematical Sciences\\
The University of the South Pacific\\
Suva, Fiji}
\email{shruti.saini@usp.ac.fj}
\and
Ansgar Fehnker
\institute{Department of Computer Science\\University of Twente\\Enschede, the Netherlands}
\email{ansgar.fehnker@utwente.nl}
}
\def\titlerunning{Evaluating SCTP using Uppaal}
\def\authorrunning{S.~Saini \& A.~Fehnker}
\begin{document}
\maketitle

\begin{abstract}
The Stream Control Transmission Protocol (SCTP) is a Transport Layer protocol that has been proposed as an alternative to the Transmission Control Protocol (TCP) for the Internet of Things (IoT). SCTP, with its four-way handshake mechanism, claims to protect the Server from a Denial-of-Service (DoS) attack by ensuring the legitimacy of the Client, which has been a known issue pertaining to the three-way handshake of TCP. This paper compares the handshakes of TCP and SCTP to discuss its shortcomings and strengths. We present an Uppaal model of the TCP three-way handshake and SCTP four-way handshake and show that SCTP is able to cope with the presence of an Illegitimate Client, while TCP fails. The results confirm that SCTP is better equipped to deal with this type of attack.
\end{abstract}

\section{Introduction}
The Internet of Things (IoT) is an emerging field envisioned to connect all physical objects to the Internet, enabling them to communicate with one another and perform tasks autonomously. The technology most commonly referenced to provide such behavior for the objects in IoT is the Radio Frequency Identification Device (RFID), which are low-powered energy-constrained devices. Due to the sheer amount of objects to be connected in IoT, there are many research challenges that need to be tackled. Some of these include its architecture, networks, applications and security. This paper looks at the Transport Layer for IoT.

In the Internet, the protocol most commonly used at the transport layer is the Transmission Control Protocol (TCP). This protocol however, has been found to not meet the requirements for IoT applications due to its high power consumption resulting from the verbose session overhead and requirements for reliability which requires packet acknowledgment \cite{Palattella2013, Sehgal2012}. In contrast, the Stream Control Transmission Protocol (SCTP) supports all features of TCP and also claims to provide protection against denial-of-service (DoS) attacks, amongst other advanced features.

We will evaluate SCTP against the SYN flooding DoS attack that is known in TCP \cite{Stalvig2007}.
SCTP, with its four-way handshake claims to protect the Server from a DoS attack by ensuring the legitimacy of the Client which has been a known issue pertaining to the three-way handshake of TCP \cite{Natarajan2009}. The SYN flooding attack and several mitigation strategies for TCP are discussed extensively in \cite{WesleyM.2007}. This paper, however, uses Uppaal, a model checking tool, to formally model and verify the basic handshake mechanisms employed by TCP and SCTP. Previous analysis of the protocol concentrated on its performance aspects \cite{boussen2009performance} and used network simulator NS-2. Different types of potential attacks were discussed in \cite{aura2004effects}, and the authors undertook a manual review of three implementations to look for vulnerabilities. This paper in contrast, built a formal model of the handshake mechanism in TCP and SCTP, to verify SCTP's resilience to a specific attack.

The next section will introduce the handshake mechanisms of TCP and SCTP.
Section \ref{sec:tcp} and \ref{sec:sctp} discuss the TCP and SCTP models, respectively. Section \ref{sec:results} will discuss the results for properties that relate to a potential SYN flooding attack.

\section{TCP and SCTP}\label{sec:tcpsctp}
Communication on the Internet is governed by the Internet Protocol (IP) Suite comprised of a set of layered protocols implemented by the host. There are five layers in the Internet Architecture, however, we will only explore protocols from the Transport Layer. The most popular and widely implemented protocol is TCP. This protocol was intended as a highly reliable host-to-host protocol between hosts in packet-switched computer communication networks, and was standardized in 1981 as RFC 793 \cite{Kozierok2005, Postel1981}.

SCTP, on the other hand, was developed to support functionalities that neither TCP nor UDP could offer. It was standardized by the Internet Engineering Task Force (IETF) in the year 2000 in the RFC 4960 \cite{Dreibholz2011, Stewart2007, Wallace2012}. The SCTP protocol has become a general purpose transport protocol with most features of TCP and a set of other features for  security, multihoming, multistreaming, mobility and partial reliability.

Both SCTP and TCP are connection oriented protocols. This means, prior to any communication between two parties, a setup procedure needs to be executed to establish a communication relationship and state. For TCP, this occurs with a three-way handshake to establish the relationship which is called a connection. For SCTP, a four-way handshake is used, and the relationship is called an association. It encompasses a broader concept than a single connection with its multihoming feature. Both TCP and SCTP use a Transmission Control Block (TCB) to hold their connection or association state information.

\subsection{Segment and Packets}
\begin{figure}[tpbh]\begin{centering}
\includegraphics[width=0.75\linewidth]{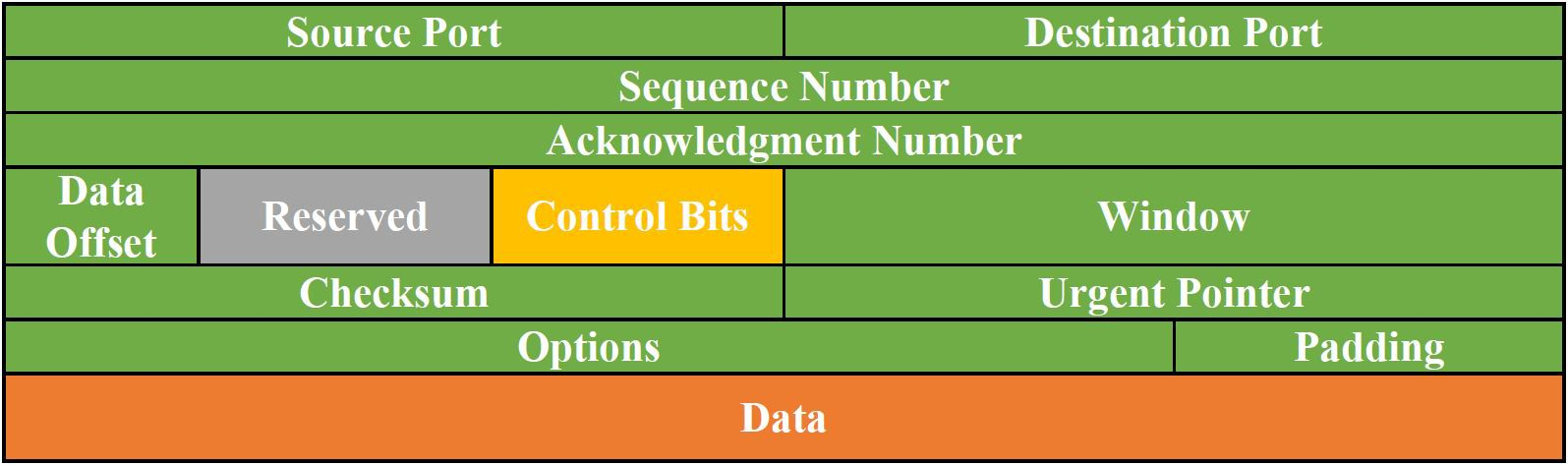}\\[-2ex]
\caption{TCP Segment format}\label{fig:tcp}\vspace{2ex}
\includegraphics[width=0.75\linewidth]{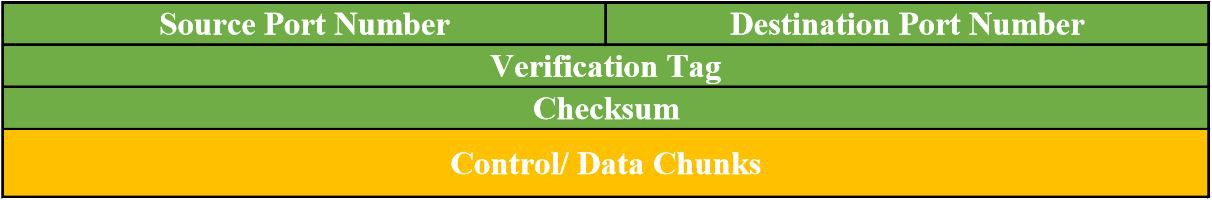}\\[-2ex]
\caption{SCTP Packet format}\label{fig:sctpheader}\vspace{2ex}
\includegraphics[width=0.75\linewidth]{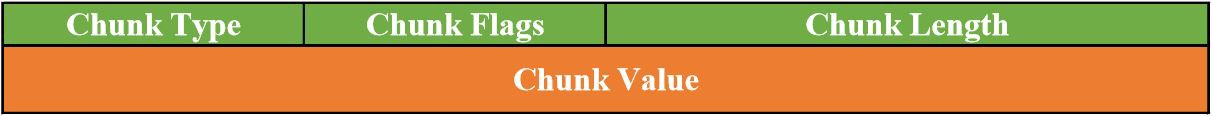}\\[-2ex]
\caption{SCTP Chunk format}\label{fig:sctpchunk}
\end{centering}
\end{figure}

Any information exchange in TCP uses segments, while SCTP uses packets. TCP encapsulates the data received from the Application Layer into a TCP segment by adding the TCP Header. The TCP Header follows the IP header, and supplies protocol specific information. Figure \ref{fig:tcp} depicts the format of the TCP header.

Every packet in SCTP consists first of a common Header and is followed by chunks containing either control information or user data. SCTP allows to bundle multiple chunks into a single packet, with some exceptions. Figure \ref{fig:sctpheader} and Figure \ref{fig:sctpchunk} depict the format of the SCTP header and the SCTP chunk.

Since every connection and association in TCP and SCTP is distinct, certain data about each needs to be maintained separately. Both TCP and SCTP, for this purpose, utilize a special data structure called the TCB that records the state of a connection and association, respectively.

For TCP, the variables stored in the TCB are the local and remote socket numbers, the state information, the security and precedence of the connection, pointers to the users send and receive buffers, pointers to the retransmit queue and to the current segment and several variables relating to the send and receive sequence numbers, amongst others.

SCTP maintains similar information about its associations like the local and remote socket numbers, the state information, a list of all local and remote transport addresses bound to the association, several variables relating to the send and receive sequence numbers and an array of structures to track the inbound and outbound streams, amongst others.

\subsection{Connections in TCP}
Communication in TCP is dictated by the establishment of a successful connection between two endpoints through the three-way handshake. The handshake includes the following steps:

\begin{enumerate}
    \item All Endpoints begin from the CLOSED state. When Endpoint A wishes to start communicating with another it performs an Active OPEN whereby it creates a TCB to store the necessary information and sends the Synchronize (SYN) segment before moving to the SYN-SENT state.

    \item An Endpoint B that wishes to receive any incoming connection request performs a Passive OPEN whereby it creates the TCB which is partially filled with unspecified foreign sockets and enters the LISTEN state. Endpoint B receives the incoming SYN segment and fills the parameters of the partially completed TCB before replying with a SYN segment itself along with an Acknowledgment (ACK) segment, collectively called the SYN+ACK. In doing so, Endpoint B allocates resources to the unestablished connection and updates the TCB to the SYN-RECEIVED state in wait for an ACK segment.

    \item Endpoint A receiving the SYN+ACK replies with an ACK segment as its final reply to establish the connection and moves to the ESTABLISHED state.

    \item Endpoint B upon receiving the final ACK segment also updates its TCB to the ESTABLISHED state and successfully establishes the connection.
\end{enumerate}
There are only two segments, SYN and ACK, involved in the three-way handshake for TCP as illustrated in Figure \ref{fig:tcphandshake}.

\subsection{Associations in SCTP}
\begin{figure}[t!]\begin{centering}\parbox[c][0.45\linewidth]{0.5\linewidth}{
\includegraphics[width=\linewidth]{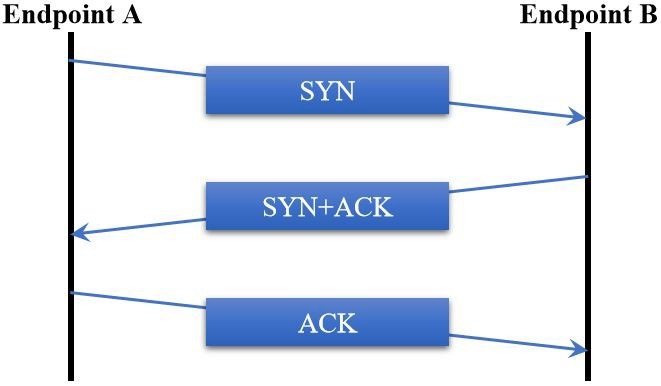} \vfill
\caption{TCP three-way handshake}\label{fig:tcphandshake}}
\parbox[c][0.45\linewidth]{0.5\linewidth}{
\includegraphics[width=\linewidth]{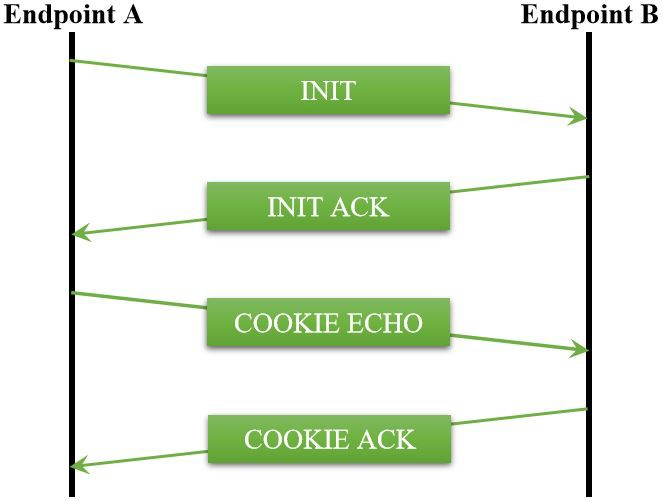} \vfill
\caption{SCTP four-way handshake}\label{fig:sctphandshake}}
\end{centering}
\end{figure}

Prior to any communication that can occur between two Endpoints in SCTP, they must first establish an association. The handshake includes the following steps:

\begin{enumerate}
    \item Initially, all Endpoints begin from the CLOSED state. When Endpoint A wishes to start communicating with another it creates a TCB to store the necessary information and sends the Initiation (INIT) chunk before moving to the COOKIE-WAIT state.

    \item	Endpoint B receiving the incoming INIT chunk creates a temporary TCB to extract a subset of information that would help recreate the TCB along with a Message Authentication Code (MAC) and a secret key which are then used to generate a cookie.  This cookie is sent as a reply with the Initiation Acknowledgment (INIT ACK) chunk. The temporary TCB is then deleted and Endpoint B remains in the CLOSED state preventing the allocation of resources for an unestablished connection.

    \item Endpoint A upon receiving the INIT ACK chunk replies by echoing the cookie back with a Cookie Echo (COOKIE ECHO) chunk and enters the COOKIE-ECHOED state.

    \item Endpoint B upon receiving the cookie back with the COOKIE ECHO chunk validates the TCB with the MAC to confirm the authenticity of the cookie. The TCB is then recreated from the information present in the cookie and a final reply is sent with the Cookie Acknowledgment (COOKIE ACK) chunk to establish the association and updates its TCB to the ESTABLISHED state. In doing so, resources are finally assigned to the association.

    \item Endpoint A receives the final COOKIE ACK chunk and moves to the ESTABLISHED state and successfully establishes the association.
\end{enumerate}
There are four chunks involved in the four-way handshake for SCTP as illustrated in Figure \ref{fig:sctphandshake}.

\section{UPPAAL Model Checker}\label{sec:uppaal}

Distributed systems are difficult to understand, design, and reason about due to their complexity and non-deterministic nature. They usually involve subtle interactions of a number of components and a high level of parallelism. This is why the correctness of these systems is difficult to ensure. Several systems and protocols have been proven not to succeed in satisfying their intended goals after they have been published \cite{Al-Bataineh2012}. One promising solution to this problem is the use of formal verification techniques such as model checking \cite{Al-Bataineh2012}.

Developed in conjunction by the Department of Computer Systems at Uppsala University, Sweden, and BRICS at Aalborg University, Denmark, UPPAAL is a tool for the modeling, simulation and verification of real-time systems \cite{Al-Bataineh2012,Behrmann2002}. The tool is appropriate for systems that can be modeled as a collection of non-deterministic processes with finite control structure and real-valued clocks, communicating through channels or shared variables \cite{Behrmann2002}.

The following Section discusses our UPPAAL implementation of the TCP and SCTP handshakes. The models implemented consider all requests to be valid and authentic. This assumption allows us to create a simulated version of the handshakes which enables the receiver to non-deterministically chose a reply to an incoming request regardless of its content and authenticity. The key focus, however, is to the handling of the first incoming request which may be susceptible to a SYN flooding attack.

\section{TCP Model}\label{sec:tcp}

According to the connection establishment process of TCP, there are five states in the three-way handshake. In the model, these are represented by constant integer variables declared as \texttt{CLOSED}, \texttt{LISTEN}, \texttt{SYN\char`_SENT}, \texttt{SYN\char`_RECEIVED} and \texttt{ESTABLISHED} states as shown in Table \ref{tab:tcpstates}. Typical application areas include real-time controllers and communication protocols. UPPAAL has been applied successfully in case studies ranging from communication protocols to multimedia applications \cite{Behrmann2004}.

\begin{table}[h]
	\begin{center}
		\begin{tabular}{l p{8cm}}
			\hline
			\textbf{State} & \textbf{Description}\\
			\hline
			\texttt{CLOSED} & No connection \\
			\texttt{LISTEN} & Waiting for any connection request \\
			\texttt{SYN\char`_SENT} & Waiting for a matching connection request after having sent a connection request \\
			\texttt{SYN\char`_RECEIVED} & Waiting for a confirming connection request acknowledgment after having both received and sent a connection request \\
			\texttt{ESTABLISHED} & An open active connection \\
		    \hline
		\end{tabular}
		\caption{Summary of States in the three-way handshake} \label{tab:tcpstates}\vspace{-3ex}
  \end{center}
\end{table}

\subsection{The Client Template}

The \emph{Legitimate Client} template for TCP as modelled in Uppaal is shown in Figure \ref{fig:tcpLegitClientTime}.  The TCP model contain a number of channels. All of these channels are broadcast channels, meaning that messages can be dropped, if the intended recipient is not able to receive the message. The descriptions for the edges of the template are as follows:

\begin{enumerate}
    \item Location LC0 is the initial location of the automaton, representing the \texttt{CLOSED} state. The \emph{Client} can perform an Active Open on the \texttt{syn} channel in order to send a connection request, set the \texttt{counter}, start the \texttt{timer} and move to location LC1.

    \item Location LC1 represents the \texttt{SYN\char`_SENT} state. Here the \emph{Client} has the following options:

    \begin{figure}[t!]
    \begin{center}
    \includegraphics[width=14cm]{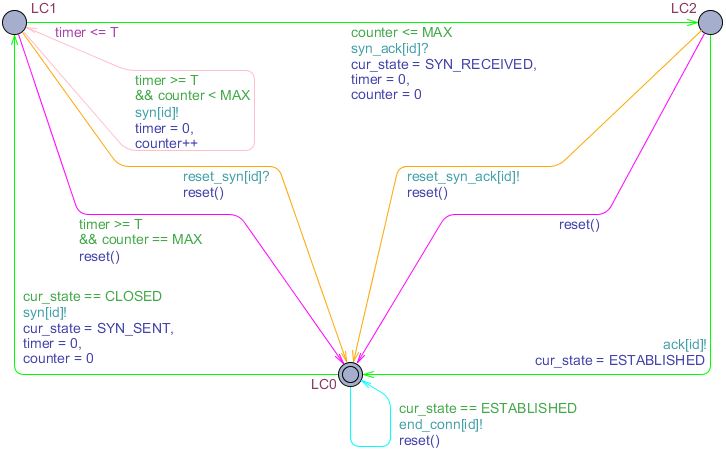}
    \end{center}
    \caption{TCP Legitimate Client template}
    \label{fig:tcpLegitClientTime}
    \end{figure}

    \begin{enumerate}
        \item \emph{Receive Acknowledgment} The \emph{Client} can receive an acknowledgment on the \texttt{syn\char`_ack} channel, move to location LC2, and update its state to \texttt{SYN\char`_RECEIVED} as long as the reply comes before the maximum number of retransmits are made and before the \texttt{timer} expires.

        \item \emph{Reset} The \emph{Client} can receive a request to reset the connection on the \texttt{reset\char`_syn} channel, move back to location LC0 and reset all state information.

        \item \emph{Retransmit} The \emph{Client} can retransmit the connection request on the \texttt{syn} channel if no reply is received in a certain time \texttt{T} and if retransmits are still allowed. The \texttt{counter} of retransmits is incremented and the \texttt{timer} restarted.

        \item \emph{Discard} The \emph{Client} can silently discard the connection request made, move back to location LC0 and reset all state information.
    \end{enumerate}

    \item Location LC2 is an intermediate state, where a \texttt{syn\char`_ack} has been received. Here the \emph{Client} has the following options:

    \begin{enumerate}
        \item \emph{Send Acknowledgment} The \emph{Client} can send the acknowledgment on the \texttt{ack} channel, move back to location LC0, and change the state to \texttt{ESTABLISHED}.

        \item \emph{Reset} The \emph{Client} can ask the \emph{Server} to reset the connection on the \texttt{reset\char`_syn\char`_ack} channel, move back to location LC0 and reset all state information.

        \item \emph{Discard} The \emph{Client} can silently discard the connection request, move back to location LC0 and reset all state information.
    \end{enumerate}

    \item If the \emph{Client} is in location L0 with current state \texttt{ESTABLISHED}, it can request to end the connection on the \texttt{end\char`_conn} channel, return to location LC0 and reset all state information.
\end{enumerate}

The \emph{Illegitimate Client} template for TCP as modelled in Uppaal has a single Location IC0 with a single self loop that keeps transmitting a connection request on the \texttt{syn} channel in an attempt to occupy all \emph{Server} resources.

\subsection{The Server Template}

Both the TCP and SCTP protocol include a TCB block. The Uppaal model of these protocols uses the same basic data structure to model the TCB block. It does not incorporate all the fields from the segments and packets or the TCB that are used during the handshake as described in \cite{Postel1981, Stewart2007}. Only the fields that are necessary for simple verification and identification of the segments and packets are kept.

The TCB is maintained for all connections and associations by the \emph{Server} Endpoint. Listing \ref{lst:commonlocalserver} shows the declaration. The constant \texttt{RESOURCES} refers to the number of possible Endpoints a connection or association can be established with, which is one less that the total number of Endpoints created.\\

\begin{minipage}[c]{0.9\linewidth}
\begin{lstlisting}[language={[GUI]Uppaal}, numbersep=8pt, numberstyle=\tiny\color{white},
columns={[l]flexible},
frameround=fftt, frame=shadowbox, rulesepcolor=\color{gray},
caption={Local Declarations of the TCB at the Server Endpoint}, label=lst:commonlocalserver]
typedef struct {
	ids peer;
	int [CLOSED, ESTABLISHED]cur_state;
} TCB;

TCB tcb[RESOURCES];
\end{lstlisting}
\end{minipage}

The \emph{Server} template includes several functions to maintain the TCB, most notably \texttt{update\_TCB} to add new connections or associations information.
A resource \texttt{i} is available if the current state \texttt{tcb[i].cur_state} is \texttt{LISTEN}. Initially all resources are available. When the ACK segment is received for an active connection request with an Endpoint the current state will be \texttt{SYN\char`_RECEIVED}.

The \emph{Server} template for TCP as modelled in Uppaal is shown in Figure \ref{fig:tcpServer}. The descriptions for the edges of the template are as follows:

\begin{figure}[t!]
\begin{center}
\includegraphics[width=\linewidth]{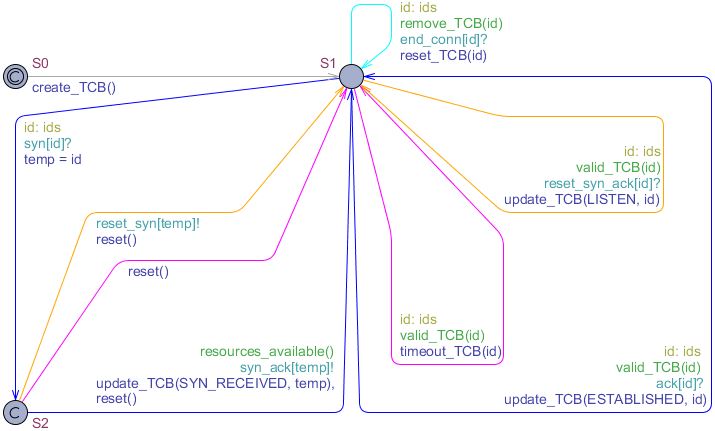}
\end{center}
\caption{TCP Server template}
\label{fig:tcpServer}
\end{figure}

\begin{enumerate}
    \item Location S0 is the initial committed location. The \emph{Server} can perform a Passive Open, create and partially initialize the TCB, move to location S1, and update its state to \texttt{LISTEN} in order to start receiving connection requests from the \emph{Client}.

    \item In location S1 the \emph{Server} Endpoint receives incoming connection requests on the \texttt{syn} channel and moves to location S2.

    \item Location S2 is an intermediate committed state. Here the \emph{Server} has the following options:

    \begin{enumerate}
        \item \emph{Send Acknowledgment} If a resource is available, the \emph{Server} can send the acknowledgment on the \texttt{syn\char`_ack} channel, move to location S1, and update its state to \texttt{SYN\char`_RECEIVED}.

        \item \emph{Reset} The \emph{Server} can ask the \emph{Client} to reset the connection on the \texttt{reset\char`_syn} channel, move back to location S1 and reset all state information.

        \item \emph{Discard} The \emph{Server} can silently discard the connection request, move back to location S1 and reset all state information.
    \end{enumerate}

    \item  In location S1 new connection requests from \emph{Clients} can be received continuously as well as requests to further or end the half-open and fully established connections:

    \begin{enumerate}
        \item \emph{Receive Acknowledgment} The \emph{Server} can receive an acknowledgment on the \texttt{ack} channel, move back to location S1, and change the state to \texttt{ESTABLISHED}.

        \item \emph{Reset} The \emph{Server} can receive a request to reset the connection on the \texttt{reset\char`_syn\char`_ack} channel, move back to location S1 and reset all state information.

        \item \emph{Time-out} The \emph{Server} can remove half-open connection requests which have timed out and reset all state information. This is a condensed representation of the time-wait and retransmission attempts to fully establish the connection before they are ceased.

        \item \emph{End Connection} The \emph{Server} can receive a request to end the connection on the \texttt{end\char`_conn} channel and reset all state information.
    \end{enumerate}
\end{enumerate}

\section{SCTP Model}\label{sec:sctp}
According to the association establishment process of SCTP, there are four states in the four-way handshake. In the model, these are represented by integer constants  \texttt{CLOSED}, \texttt{COOKIE\char`_WAIT}, \texttt{COOKIE\char`_ECHOED} and \texttt{ESTABLISHED}, as shown in
 Table \ref{tab:sctpstates}.

\begin{table}[h]
	\begin{center}
		\begin{tabular}{l p{8cm}}
			\hline
			\textbf{State} & \textbf{Description}\\
			\hline
			\texttt{CLOSED} & No association \\
			\texttt{COOKIE\char`_WAIT} & Waiting for a confirming association request acknowledgment with a cookie after having sent an association request \\
			\texttt{COOKIE\char`_ECHOED} & Waiting for a confirming association request acknowledgment after having both received and sent the cookie back \\
			\texttt{ESTABLISHED} & An open active association \\
		    \hline
		\end{tabular}
		\caption{Summary of States in the four-way handshake}
		\label{tab:sctpstates}
  \end{center}
\end{table}

\subsection{The Client Template}

The \emph{Legitimate Client} template for SCTP as modelled in Uppaal is shown in Figure \ref{fig:sctpLegitClientTime}. The descriptions for the edges of the template are as follows:

\begin{figure}[b!]
\begin{center}
\includegraphics[width=14cm]{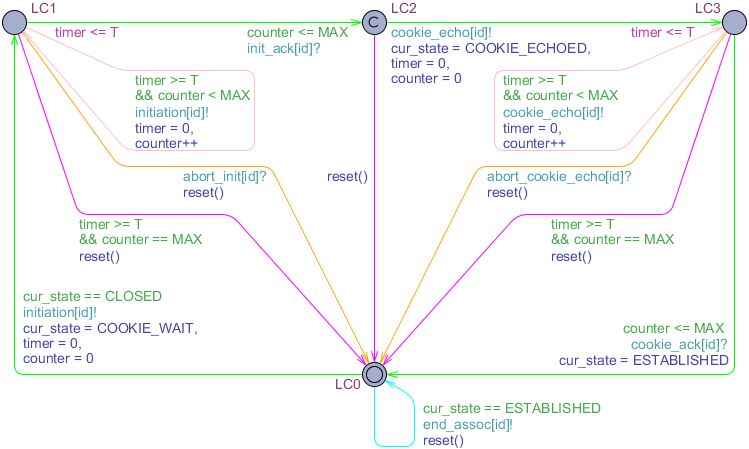}
\end{center}
\caption{SCTP Legitimate Client template}
\label{fig:sctpLegitClientTime}
\end{figure}

\begin{enumerate}
    \item Location LC0 is the initial location of the automaton, representing the \texttt{CLOSED} state. The \emph{Client} can send an association request over the \texttt{initiation} channel, set the \texttt{counter}, start the \texttt{timer} and move to location LC1.

    \item Location LC1 represents the \texttt{COOKIE\char`_WAIT} state. Here the \emph{Client} has the following options:

    \begin{enumerate}
        \item \emph{Receive Acknowledgment} The \emph{Client} can receive an acknowledgment on the \texttt{init\char`_ack} channel, move to location LC2, as long as the reply comes before the maximum number of retransmits are made and before the \texttt{timer} expires.

        \item \emph{Abort} The \emph{Client} can receive a request to abort the association over the \texttt{abort\char`_init} channel, move back to location LC0 and reset all state information.

        \item \emph{Retransmit} The \emph{Client} can retransmit the association request over the \texttt{initiation} channel if no reply is received in a certain time \texttt{T} and if retransmits are still allowed. The \texttt{counter} of retransmits is incremented and the \texttt{timer} restarted.

        \item \emph{Discard} The \emph{Client} can silently discard the association request, move back to location LC0 and reset all state information.
    \end{enumerate}

    \item Location LC2 is an intermediate committed state, where a \texttt{init\char`_ack} has been received. Here the \emph{Client} has the following options:

    \begin{enumerate}
        \item \emph{Send Acknowledgment} The \emph{Client} can send the acknowledgment on the \texttt{cookie\char`_echo} channel, set the \texttt{counter}, start the \texttt{timer}, and move to location LC3.

        \item \emph{Discard} The \emph{Client} can silently discard the association request, move back to location LC0 and reset all state information.
    \end{enumerate}

    \item Location LC3 represents the \texttt{COOKIE\char`_ECHOED} state where the cookie has been echoed back by the \emph{Client}. The transitions available here are similar to those in location LC1.

    \item If the \emph{Client} is in location LC0 with current state \texttt{ESTABLISHED}, it can request to end the association over the \texttt{end\char`_assoc} channel, return to location LC0 and reset all state information.
\end{enumerate}

The \emph{Illegitimate Client} template for SCTP is similar to the model for TCP. It has only one location, and a self loop that keeps transmitting an association request over the \texttt{initiation} channel in an attempt to occupy all \emph{Server} resources.

\subsection{The Server Template}

The \emph{Server} template uses the same TCB block as the TCP Server template. The Uppaal model is shown in Figure \ref{fig:sctpServer}. The descriptions for the edges of the template are as follows:

\begin{figure}[b!]
\begin{center}
\includegraphics[width=14cm]{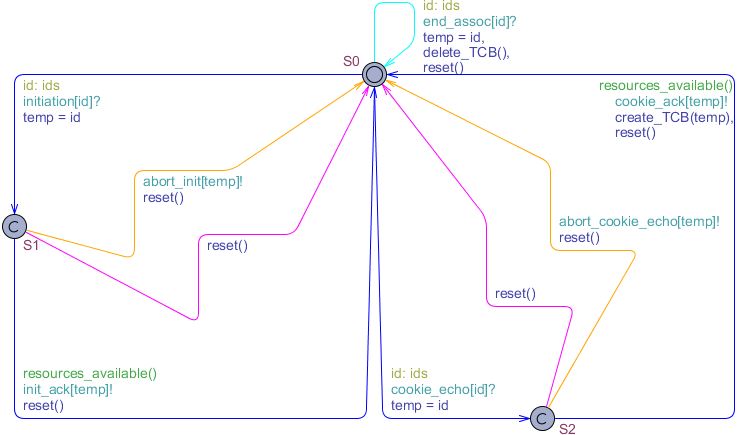}
\end{center}
\caption{SCTP Server template}
\label{fig:sctpServer}
\end{figure}

\begin{enumerate}
    \item Location S0 is the initial location of the \emph{Server} Endpoint which receives incoming association requests on the \texttt{initiation} channel and moves to location S1.

    \item Location S1 is an intermediate committed state. Here the \emph{Server} has the following options:

    \begin{enumerate}
        \item \emph{Send Acknowledgment} If a resource is available, the \emph{Server} can send the acknowledgment on the \texttt{init\char`_ack}, move back to location S0 and keep no state information.

        \item \emph{Abort} The \emph{Server} can ask the \emph{Client} to abort the association on the \texttt{abort\char`_init} channel, move back to location S0 and keep no state information.

        \item \emph{Discard} The \emph{Server} can silently discard the association request, move back to location S0 and keep no state information.
    \end{enumerate}

    \item In location S0 new association requests from \emph{Clients} can be received continuously as well as requests to further or end the associations:

    \begin{enumerate}
        \item \emph{Receive Acknowledgment} The \emph{Server} can receive an acknowledgment on the \texttt{cookie\char`_echo} channel and move to location S2 which is an intermediate committed state. The transitions available here are similar to those in location S1.

        \item \emph{End Association} The \emph{Server} can receive a request to end the association over the \texttt{end\char`_assoc} channel and reset all state information.
    \end{enumerate}

\end{enumerate}

\section{Verification Results}\label{sec:results}

We considered a number of properties to explore the correctness of TCP and SCTP. This section discusses two properties that highlight differences between TCP and SCTP, especially weaknesses in TCP.

The property in Listing 2 checks if in all states along all paths it holds that if a \emph{Legitimate Client} is in the \texttt{ESTABLISHED} state, then a  corresponding \emph{Server} \texttt{tcb} resource is also in the \texttt{ESTABLISHED} state. If a \emph{Legitimate Client} has an active connection or association then there must be a corresponding \emph{Server} \texttt{tcb} resource in the fully-established state.

However, it appears that the TCP model does not produce the desired results, and examination confirms that this a known problem with TCP. The TCP model allows for half-open connections where the \texttt{ack} is not received by the \emph{Server}. Failure to satisfy the property in Listing \ref{lst:p8} shows that the TCP model allows a \emph{Legitimate Client} to reach the \texttt{ESTABLISHED} state while the \emph{Server} remains in some other arbitrary state.\\

\begin{minipage}[c]{0.9\linewidth}
\begin{lstlisting}[language={[GUI]Uppaal}, numbersep=8pt, numberstyle=\tiny\color{white},
columns={[l]flexible},
frameround=fftt, frame=shadowbox, rulesepcolor=\color{gray},
caption={For any active connection should involve a server in the fully-established state.}, label=lst:p8]
A[] forall (i:ids) (
  Legit_Client(i).cur_state == ESTABLISHED imply
    exists (j: int[0,(RESOURCES-1)])(
      Server.tcb[j].peer == i and Server.tcb[j].cur_state == ESTABLISHED
    )
)
\end{lstlisting}
\end{minipage}

Half-open connections are a known problem of TCP  and they occurs due to a number of reasons in a real world application. A slow or lossy network, for example, can lead to the \texttt{ack} not being received by the \emph{Server} in time, resulting in the \emph{Client} assuming the connection was successfully established while the \emph{Server} may remain in an arbitrary state.

This behavior of TCP is known to be exploited for  SYN flooding attack. The attacker merely attempts to send enough \texttt{syn} requests, engaging its resources. Once a backlog of bogus half-open connections are established, the \emph{Server} is not able to process requests from \emph{Legitimate Clients}.

SCTP, in comparison, does not allow these half-open associations. Every association request in SCTP is replied to without allocating it any \texttt{tcb} resources. One can argue that the cookie may fall susceptible to the SYN flooding attack, however the purpose of the model was not to authenticate received chunks but to simulate all potential replies. The model provides the \emph{Server}, as per its specification, the ability to \emph{Send Acknowledgment}, \emph{Abort} or \emph{Discard} incoming cookies same as the first association request. The authentication of the cookie is assumed but through the implementation we are able to observe the cookie mechanism of SCTP successfully keep its resources free until the received cookie is authenticated by the \emph{Server}.

The property in Listing 3 checks if any resource has been  allocated to the same \emph{Client}. For SCTP it is defined as follows:\\

\begin{minipage}[c]{0.9\linewidth}
\begin{lstlisting}[language={[GUI]Uppaal}, numbersep=8pt, numberstyle=\tiny\color{white},
columns={[l]flexible},
frameround=fftt, frame=shadowbox, rulesepcolor=\color{gray},
caption={For SCTP we check if the current state is in \texttt{SYN\_RECEIVED}, instead of not equal to \texttt{Closed}.}, label=lst:p101]
E<> exists (i:ids) (
   forall (j: int[0,(RESOURCES-1)])(
      Server.tcb[j].peer == i and  Server.tcb[j].cur_state != CLOSED
   )
)
\end{lstlisting}
\end{minipage}

If the model satisfies this property, it means that a resource has been successfully hogged by an \emph{Illegitimate Client}. TCP satisfies this property, and thus fails to prevent hogging of resources. It allows the \emph{Illegitimate Clients} to successfully occupy all \emph{Server} \texttt{tcb} resources while attempting to establish a connection. In contrast, all SCTP models promptly replied to the \emph{Illegitimate Client} association request without allocating it any \emph{Server} \texttt{tcb} resource. This keeps the \emph{Server} \texttt{tcb} resources free which prevents it from a DoS since the resources are not tied up with \emph{Illegitimate Client} requests.

Although, the \emph{Illegitimate Clients} are not able to establish a connection in TCP, the backlog of these half-open connections allows for a DoS like the SYN flooding attack. From this, we are able to formally verify and observe how SCTP using its cookie authentication is able to successfully prevent DoS attacks, as claimed.

\section{Conclusion}
This paper analysed network issues at the Transport Layer and confirmed that the TCP protocol does not provide basic security against DoS attacks on the IoT enabled devices. To clearly understand the differences between the two protocols, we took a detailed look into their handshake mechanisms which can be vulnerable to a DoS attack like the SYN flooding attack in TCP. A model-checker Uppaal was used to formally test the protocol's handshake mechanisms and to test SCTP's claims and TCP's vulnerability. We were able to confirm TCP's susceptibility to DoS attacks, as well as  SCTP's ability in preventing it.

In conclusion, we were able to successfully evaluate SCTP to check for its applicability to IoT in comparison with TCP. We can conclude that the handshake mechanism of SCTP does in fact provide protection against DoS attacks, fulfilling a security requirement of IoT.

\bibliographystyle{eptcs}
\bibliography{generic}
\end{document}